\documentclass[12pt]{article} 
\usepackage{graphicx}
\usepackage{epsfig}
\usepackage{amsbsy}

\begin{document}
\baselineskip 10mm

\centerline{\large \bf Violation of the Equipartition Theorem}
\centerline{\large \bf for Thermally Insulated Clusters of Atoms with Different Masses}

\vskip 6mm

\centerline{L. A. Openov$^{*}$ and A. I. Podlivaev}

\vskip 4mm

\centerline{\it Moscow Engineering Physics Institute (State
University), 115409 Moscow, Russia}

\vskip 2mm

$^{*}$ E-mail: LAOpenov@mephi.ru

\vskip 8mm

\centerline{\bf ABSTRACT}

An expression is derived for calculating microcanonical-ensemble averages of the kinetic energies of atoms of different types in clusters isolated from the environment. This expression is a natural generalization of the solution to the problem of hard spheres with different masses to a system with a many-particle interatomic interaction potential. The dynamics of a C$_8$H$_8$ cubane is simulated numerically. The data on the numerical simulation confirm the validity of the results obtained.

\newpage

Statistical averages of physical quantities can be determined using ensembles of different types, such as microcanonical, canonical, large canonical, and other ensembles. In this case, the results obtained for macroscopic systems with numbers of particles $N\gg 1$ differ
by a negligibly small value of the order of $O(1/N)$ [1]. Therefore, frequently (although not necessarily), the choice of a particular ensemble is determined by the convenience of calculations. In numerical simulation of the dynamics of finite systems, the number of particles
and their total energy $E$ (as well as the volume $V$ occupied by these particles) often remain unchanged during the evolution, which corresponds to a microcanonical $NVE$ ensemble [1,2]. In many models, the total momentum {\bf P} is also an integral of motion ($NVE{\bf P}$ ensemble [3-5]. From the physical standpoint, this statement of the problem corresponds to systems that do not interact with the environment (for example, excited clusters in a gas phase [6-13]).

If particles are classical and the energy of the interparticle interaction depends only on their relative coordinates (rather than on velocities), the kinetic energy per degree of freedom in a canonical ensemble (i.e., an
ensemble of systems that are in thermal equilibrium with a reservoir at a temperature $T$) is equal to $\frac{1}{2}k_BT$, where $k_B$ is the Boltzmann constant. This is known as the equipartition theorem. Correspondingly, the kinetic energy of each particle (irrespective of its mass) is $\langle \varepsilon_{kin}^i\rangle_c=\frac{3}{2}k_BT$, where $i=1-N$ is the particle number (hereafter, it will be assumed that the system is three-dimensional) and the total kinetic energy is represented in the form
\begin{equation}
\langle E_{kin}\rangle_c=\frac{3}{2}k_BTN  ,
\label{Ekin_c}
\end{equation}
where $\langle ... \rangle_c$ indicates the canonical-ensemble average.

The relationship similar to expression (1) is valid for a microcanonical
$NVE$ ensemble [2]; that is,
\begin{equation}
\langle E_{kin}\rangle_{NVE}=\frac{3}{2}k_BT_mN  ,
\label{Ekin_NVE}
\end{equation}
where $T_m$ is the microcanonical temperature defined by the formula
\begin{equation}
\frac{1}{T_m}=\left[\frac{\partial S}{\partial E}\right]_V  .
\label{Tm}
\end{equation}
Here $S=k_B\ln\Omega$ is the entropy and $\Omega(N,V,E)$ is the phase space volume proportional to the number of states with energies lower than or equal to $E$. Note that the definition of the entropy $S$ through the logarithm of the density of states $\omega=\partial\Omega/\partial E$ leads to another relationship between the quantities$\langle E_{kin} \rangle_{NVE}$ and $T_m$, which transforms into expression (2) at $N\rightarrow\infty$ (for more detail, see [2]). Despite the formal similarity of expressions (1) and (2), the equipartition in the $NVE$
ensemble can be considered only "integrally" because it does not
follow from relationship (2) that the kinetic energy $\langle\varepsilon_{kin}^i\rangle_{NVE}$ averaged over the
$NVE$ ensemble is $\frac{3}{2}k_BT_m$ for {\it each} particle .

If the total momentum of the system is ${\bf P}=0$, relationship (2) takes the form [3-5]
\begin{equation}
\langle E_{kin}\rangle_{NVE{\bf P}}=\frac{3}{2}k_BT_m(N-1)
\label{Ekin_NVEP}
\end{equation}
because the number of degrees of freedom decreases by three [note that the microcanonical temperatures $T_m$ in expressions (2) and (4) are different, since the additional condition ${\bf P}=0$ results in a change in the phase
space volume $\Omega$]. When, in addition, the angular momentum is ${\bf J}=0$, we have
\begin{equation}
\langle E_{kin}\rangle_{NVE{\bf PJ}}=\frac{3}{2}k_BT_m(N-2)  ,
\label{Ekin_NVEPJ}
\end{equation}
because the number of degrees of freedom decreases by
six. For ergodic systems, the microcanonical-ensemble
average is equal to the average $\langle ... \rangle_t$ over the time of the
evolution of one individual system. In this case, the microcanonical temperature $T_m$ coincides with the "dynamical temperature" $T_d$ that, at
${\bf P}=0$ and ${\bf J}=0$, is determined from the formula [14]
\begin{equation}
\langle E_{kin}\rangle_t=\frac{3}{2}k_BT_d(N-2)  .
\label{Ekin_t}
\end{equation}
It should be noted that the use of periodic boundary conditions in numerical calculations (as a rule, in order to weaken the effect of finite sizes) is equivalent to the isolation of the system from the action of external forces. This leads to the conservation of the total momentum (but not the
total angular momentum) [4].

Recently, Shirts et al. [15] succeeded in deriving the exact analytical relationships for the one-particle energy, momentum, and velocity distribution functions for a model system consisting of $N$ hard spheres with $E=$ const. In particular, it was demonstrated that the quantity
$\langle \varepsilon_{kin}^i\rangle_{NVE}=\frac{3}{2}k_BT_m$ in the
$NVE$ ensemble is identical for all atoms, whereas the equipartition theorem
in the $NVE{\bf P}$ ensemble with particles of different
masses $m_i$ is violated; that is,
\begin{equation}
\frac{\langle \varepsilon_{kin}^a\rangle_{NVE{\bf P}}}{\langle \varepsilon_{kin}^b\rangle_{NVE{\bf P}}}=\frac{M_{tot}-m_a}{M_{tot}-m_b}~  ,
\label{Shirts}
\end{equation}
where $M_{tot}=\sum_{i=1}^{N}m_i$ is the total mass of the system.
In small systems, the ratio $\langle\varepsilon_{kin}^a\rangle_{NVE{\bf P}}/\langle \varepsilon_{kin}^b\rangle_{NVE{\bf P}}$ can differ substantially from unity. This was confirmed in [15] using numerical calculations with periodic boundary conditions. Physically, the above effect is explained by the fact that the condition ${\bf P}=$ const differently
affects the energy distribution of particles with different masses (for example, the maximum energy of one particle turns out to be dependent on the particle mass [15]).

Shirts et al. [15] noted that their results can (possibly, with some corrections) appear to be valid for systems with realistic interatomic interaction potentials. In the present work, we first prove the equality of average kinetic energies of interacting particles in an $NVE$ ensemble and derive formula (7) for interacting particles in an $NVE{\bf P}$ ensemble. Second, we demonstrate how this formula is changed when both the momentum
and the angular momentum are conserved in the system. Third, we perform the numerical simulation of the dynamics of a C$_8$H$_8$ cubane and show that the simulation results are in good agreement with the derived expression. Finally, it is demonstrated using the C$_8$H$_8$ cubane as an example that the equipartition theorem holds true in the presence of the heat exchange with the environment.

The average of the physical quantity $A({\bf p},{\bf r})$ over an $NVE$ ensemble can be represented in the form [2]
\begin{equation}
\langle A\rangle_{NVE}=\frac{\int\prod_{i=1}^{N}d{\bf p}_id{\bf r}_i A({\bf p},{\bf r})\delta(E-H({\bf p},{\bf r}))}
{\int\prod_{i=1}^{N}d{\bf p}_id{\bf r}_i\delta(E-H({\bf p},{\bf r}))}~,
\label{A(E)}
\end{equation}
where ${\bf p}=\{{\bf p}_i\}$ and ${\bf r}=\{{\bf r}_i\}$ are sets of momenta and coordinates of all particles, respectively, and
$H({\bf p},{\bf r})=\sum_{i=1}^{N}{\bf p}_i^2/2m_i+U({\bf r})$ is the Hamiltonian of the system. Let us determine the relation of the average
$\langle A\rangle_{NVE}$ to the canonical-ensemble average $A(T)$, which is given by the formula\begin{equation}
A(T)\equiv\langle A\rangle_c=\frac{\int\prod_{i=1}^{N}d{\bf p}_id{\bf r}_i \exp\left(-\frac{H({\bf p},{\bf r})}{k_BT}\right)A({\bf p},{\bf r})}
{\int\prod_{i=1}^{N}d{\bf p}_id{\bf r}_i \exp\left(-\frac{H({\bf p},{\bf r})}{k_BT}\right)}~.
\label{A(T)}
\end{equation}
By multiplying the integrands in the numerator and the
denominator of relationship (9) by
$1=\int dE\delta(E-H({\bf p},{\bf r}))$, changing the order of integration over the energy $E$ and the phase space, and using formula (8), we obtain
\begin{equation}
A(T)=\frac{\int dE\exp\left(-\frac{E}{k_BT}\right)\omega(E)\langle A\rangle_{NVE}}{\int dE\exp\left(-\frac{E}{k_BT}\right)\omega(E)}~,
\label{A(T)2}
\end{equation}
where $\omega (E)$ is the denominator of formula (8), i.e., the density of states of the system. According to expression (10), the ratio between the averages $A(T)$ and $B(T)$ is written in the following form:
\begin{equation}
\frac{A(T)}{B(T)}=\frac{\int dE\exp\left(-\frac{E}{k_BT}\right)\omega(E)\langle A\rangle_{NVE}}{\int dE\exp\left(-\frac{E}{k_BT}\right)\omega(E)\langle B\rangle_{NVE}}~.
\label{A(T)B(T)}
\end{equation}
By setting $A({\bf p},{\bf r})=\varepsilon_{kin}^a={\bf p}_a^2/2m_a$ and
$B({\bf p},{\bf r})=\varepsilon_{kin}^b={\bf p}_b^2/2m_b$ and taking into account that, for the canonical ensemble, the equality
$\langle \varepsilon_{kin}^a\rangle_c=\langle \varepsilon_{kin}^b\rangle_c=\frac{3}{2}k_BT$ is true irrespective of the particle mass, we find from relation (11) that
\begin{equation}
\frac{\int dE\exp\left(-\frac{E}{k_BT}\right)\omega(E)\langle\varepsilon_{kin}^a\rangle_{NVE}}
{\int dE\exp\left(-\frac{E}{k_BT}\right)\omega(E)\langle\varepsilon_{kin}^b\rangle_{NVE}}=1~.
\label{ekin_ab}
\end{equation}
The right-hand side of this equality does not depend on the temperature $T$. This is possible only when the equality $\langle\varepsilon_{kin}^a\rangle_{NVE}=\gamma\langle\varepsilon_{kin}^b\rangle_{NVE}$ is satisfied, where $\gamma$ is a constant independent of the energy $E$. It follows from formula (12) that $\gamma=1$, i. e., $\langle\varepsilon_{kin}^a\rangle_{NVE}=\langle\varepsilon_{kin}^b\rangle_{NVE}$ at all energies $E$. Therefore, the kinetic energy average over the $NVE$ ensemble is identical for all particles and, according to relationship (2), can be represented in the form $\langle\varepsilon_{kin}^i\rangle_{NVE}=\frac{3}{2}k_BT_m$.

For an $NVE{\bf P}$ ensemble, the average of the quantity  $A({\bf p},{\bf r})$ is written in the form [3]
\begin{equation}
\langle A\rangle_{NVE{\bf P}}=\frac{\int\prod_{i=1}^{N}d{\bf
p}_id{\bf r}_i A({\bf p},{\bf r})\delta(E-H({\bf p},{\bf
r}))\delta\left({\bf P}-\sum_{i=1}^{N}{\bf p}_i\right)}
{\int\prod_{i=1}^{N}d{\bf p}_id{\bf r}_i\delta(E-H({\bf p},{\bf
r}))\delta\left({\bf P}-\sum_{i=1}^{N} {\bf p}_i \right) } ~.
\label{AP(E)}
\end{equation}
The canonical-ensemble average $A_{{\bf P}}(T)$ in the presence of the constraint on the total momentum is represented as follows:
\begin{equation}
A_{{\bf P}}(T)\equiv\langle A\rangle_{c,{\bf P}}=\frac{\int\prod_{i=1}^{N}d{\bf p}_id{\bf r}_i \exp\left(-\frac{H({\bf p},{\bf r})}{k_BT}\right)A({\bf p},{\bf r})\delta\left({\bf P}-\sum_{i=1}^{N}{\bf p}_i\right)}
{\int\prod_{i=1}^{N}d{\bf p}_id{\bf r}_i \exp\left(-\frac{H({\bf p},{\bf r})}{k_BT}\right)\delta\left({\bf P}-\sum_{i=1}^{N}{\bf p}_i\right)}~.
\label{AP(T)}
\end{equation}
By using the same procedure as for the $NVE$ ensemble, we obtain:
\begin{equation}
A_{{\bf P}}(T)=\frac{\int dE\exp\left(-\frac{E}{k_BT}\right)\omega_{{\bf P}}(E)\langle A\rangle_{NVE{\bf P}}}{\int dE\exp\left(-\frac{E}{k_BT}\right)\omega_{{\bf P}}(E)}~,
\label{AP(T)2}
\end{equation}
where $\omega_{{\bf P}}(E)$ is the denominator in formula (13).

As a result, the ratio between the energies $\langle\varepsilon_{kin}^a\rangle_{c,{\bf P}}$ and $\langle\varepsilon_{kin}^b\rangle_{c,{\bf P}}$ is given by the formula
\begin{equation}
\frac{\langle\varepsilon_{kin}^a\rangle_{c,{\bf P}}}{\langle\varepsilon_{kin}^b\rangle_{c,{\bf P}}}=
\frac{\int dE\exp\left(-\frac{E}{k_BT}\right)\omega_{{\bf P}}(E)\langle\varepsilon_{kin}^a\rangle_{NVE{\bf P}}}{\int dE\exp\left(-\frac{E}{k_BT}\right)\omega_{{\bf P}}(E)\langle\varepsilon_{kin}^b\rangle_{NVE{\bf P}}}~.
\label{eaeb}
\end{equation}
In order to calculate the energy $\langle\varepsilon_{kin}^i\rangle_{c,{\bf P}}$, we change over in formula (14) to new Jacobi momentum variables
[4, 16]:
\begin{equation}
{\bf P}_k=\frac{m_{k+1}}{M_{k+1}}\sum_{i=1}^{k}{\bf p}_i-\frac{M_k}{M_{k+1}}{\bf p}_{k+1},~k\leq N-1 ,
\label{Pk}
\end{equation}
where $M_k=\sum_{i=1}^{k}m_i$ and ${\bf P}_N=\sum_{i=1}^{N}{\bf p}_i$. Since the total kinetic energy is represented as
$E_{kin}=\sum_{i=1}^{N}{\bf p}_i^2/2m_i=\sum_{k=1}^{N}{\bf P}_k^2/2\mu_k$, where $\mu_k=m_{k+1}M_k/M_{k+1}$ for $k\leq N-1$ and $\mu_N=M_{tot}$, and the Jacobian of the transformation from $\{{\bf p}_i\}$ to $\{{\bf P}_k\}$ is equal to unity [4, 16], from expression (14) at ${\bf P}=0$ we derive the following relationship for the $N$th particle:
\begin{equation}
\langle\varepsilon_{kin}^N\rangle_{c,{\bf P}}=
\frac{3}{2}k_BT\frac{\mu_{N-1}}{m_N}=\frac{3}{2}k_BT\left(1-\frac{m_N}{M_{tot}}\right)~,
\label{eN}
\end{equation}
where we used the expression ${\bf p}_N=-{\bf P}_{N-1}$, which follows
from formula (17). Since the Jacobi momenta depend explicitly on the atomic numbering and any particle can be chosen as the $N$th particle, the following equality is satisfied:
\begin{equation}
\langle\varepsilon_{kin}^i\rangle_{c,{\bf
P}}=\frac{3}{2}k_BT\left(1-\frac{m_i}{M_{tot}}\right), ~  i=1-N~.
\label{ei}
\end{equation}
From relationships (16) and (19), we have
\begin{equation}
\frac{\int dE\exp\left(-\frac{E}{k_BT}\right)\omega_{{\bf P}}(E)\langle\varepsilon_{kin}^a\rangle_{NVE{\bf P}}}{\int dE\exp\left(-\frac{E}{k_BT}\right)\omega_{{\bf P}}(E)\langle\varepsilon_{kin}^b\rangle_{NVE{\bf P}}}=
\frac{M_{tot}-m_a}{M_{tot}-m_b}~.
\label{eaeb2}
\end{equation}
Since the right-hand side of this equality does not depend on the temperature $T$, we obtain formula (7) for the ratio $\langle\varepsilon_{kin}^a\rangle_{NVE{\bf
P}}/\langle\varepsilon_{kin}^b\rangle_{NVE{\bf P}}$. This means that the kinetic energy in the $NVE{\bf P}$ ensemble for an arbitrary
interparticle interaction is distributed among particles in the same manner as in the model problem of hard spheres [15]. In this case, from relationship (4), we derive
\begin{equation}
\langle\varepsilon_{kin}^i\rangle_{NVE{\bf
P}}=\frac{3}{2}k_BT_m\left(1-\frac{m_i}{M_{tot}}\right)~.
\label{eaeb3}
\end{equation}

The results obtained allow for the simple physical interpretation. Since the total momentum in the $NVE$ ensemble is not constant, the energy $\langle\frac{M_{tot}{\bf V}^2}{2}\rangle_{NVE}=\frac{3}{2}k_BT_m$,
(where ${\bf  V}$ is the velocity of the center of mass) corresponds
to three degrees of freedom associated with the translational motion of the system as a whole. The contribution of the $i$th particle to the energy $\langle\frac{M_{tot}{\bf V}^2}{2}\rangle_{NVE}$ is proportional to the particle mass and, hence, is $\frac{3}{2}k_BT_m\frac{m_i}{M_{tot}}$. At
${\bf P}=0$ (i. e., at ${\bf V}=0$) the average kinetic energy of the $i$th particle decreases by the above value and, therefore, is given by the expression $\frac{3}{2}k_BT_m(1-\frac{m_i}{M_{tot}})$.

By performing a similar analysis, it is possible to determine the average kinetic energy of the $i$th particle in the case of the additional constraint on the angular momentum ${\bf J}=0$. We assume that particles (atoms) form a cluster in which each particle executes vibrations with respect to its equilibrium position. At a small amplitude of these vibrations, to each of the three degrees of freedom associated with the rotation of the cluster as a whole in the $NVE$ ensemble there approximately corresponds the energy
$\langle\frac{I_n^{tot}\Omega_n^2}{2}\rangle_{NVE}=\frac{1}{2}k_BT_m$, where $I_n^{tot}$ are principal moments of inertia of the cluster with respect to its center of mass in equilibrium and $\Omega_n$ are components of the angular velocity ($n=1-3$). The contribution of the $i$th particle to the energy $\langle\frac{I_n^{tot}\Omega_n^2}{2}\rangle_{NVE}$ is proportional to its moment of inertia $I_n^i$ and, hence, is $\frac{1}{2}k_BT_m\frac{I_n^i}{I_n^{tot}}$. As a result, at ${\bf J}=0$ (i. e., at ${\bf \Omega}=0$) the average kinetic energy of the $i$th particle
decreases by $\sum_{n=1}^3\frac{1}{2}k_BT_m\frac{I_n^i}{I_n^{tot}}$ and (with allowance made for ${\bf P}=0$) is represented by the relationship
\begin{equation}
\langle\varepsilon_{kin}^i\rangle_{NVE{\bf PJ}}=\frac{3}{2}k_BT_m\left(1-\frac{m_i}{M_{tot}}-
\frac{1}{3}\sum_{n=1}^3\frac{I_n^i}{I_n^{tot}}\right)~.
\label{eaeb4}
\end{equation}
Although relationship (22) is derived not rigorously but only for states close to equilibrium states (in which the moments of inertia $I_n^i$ can be approximately considered constant during the evolution), this relationship is in the excellent agreement with the molecular dynamics
data.

The C$_8$H$_8$ cubane [17] was chosen for the numerical calculation of the distribution of the kinetic energy among atoms of different types. In this metastable cluster, the carbon atoms occupy cube vertices and the hydrogen atoms are located at extensions of the main diagonals (Fig. 1). We used the nonorthogonal tight-binding model for hydrocarbon systems [18], which was modified in our earlier work [19] in order to provide better agreement between theoretical and experimental values of the binding energies and interatomic distances in different molecules and clusters C$_n$H$_m$. For
the bond lengths in the cubane, this model leads to $l_{\mathrm{CC}}=1.5696$ \AA~ and $l_{\mathrm{CH}}=1.0823$ \AA , which are close to the experimental bond lengths, to 1.571 \AA~ and 1.097 \AA , respectively.

At the initial instant of time, random velocities and random displacements were assigned to each atom, so that the momentum and angular momentum of the
whole system were equal to zero. Then, the classical Newton equations were numerically solved with the time step $t_0=2.72\cdot 10^{-16}$ c. The total cluster energy $E=E_{pot}+E_{kin}$ remained constant in the course of simulation. The excitation energy $E_{ex}$ equal to the difference between the total energy and the energy of the cubane in the equilibrium state was chosen not very high (in order to ensure the cluster lifetime long enough to collect sufficiently large statistics before decay of the cluster) and not very low (in order for the autocorrelation times determined by the anharmonicity of vibrations be as short as possible). The determined optimum
excitation energy $E_{ex}$ corresponds to a dynamical temperature $T_d\sim 1000$ K [see relationship (6)]. In this case, $\langle E_{kin}\rangle_t\approx 0.492 E_{ex}\neq 0.5 E_{ex}$. This means that the anharmonicity effects, while weakly pronounced, occur in the system.

Figure 2 shows two characteristic calculated dependences of the ratio between the evolution time-averaged kinetic energies (i.e., the kinetic energies averaged along the trajectory of cluster motion in the phase space) of the carbon and hydrogen subsystems on the number of molecular dynamics steps $N$. The statistical processing of the results leads to the ratios $\langle E_{kin,\mathrm{C}}\rangle_t/\langle
E_{kin,\mathrm{H}}\rangle_t=0.8153 \pm0.0011$ and $0.8161 \pm0.0008$
at dynamical temperatures $T_d\approx 900$ K and 1000 K, respectively, which, within the limits of error, coincide with the ratio
$\langle E_{kin,\mathrm{C}}^i\rangle_{NVE{\bf PJ}}/ \langle
E_{kin,\mathrm{H}}^i\rangle_{NVE{\bf PJ}}=0.8155$, calculated from relationship (22) with due regard for the cubane parameters
$M_{tot}=8m_{\mathrm{C}}+8m_{\mathrm{H}}$,
$I_n^{tot}=8I_n^{\mathrm{C}}+8I_n^{\mathrm{H}}$,
$I_n^{\mathrm{H}}/I_n^{\mathrm{C}}=\frac
{m_{\mathrm{H}}}{m_{\mathrm{C}}}(1+\frac{2}{\sqrt{3}}\frac{l_{\mathrm{CH}}}{l_{\mathrm{CC}}})^2$, and $m_{\mathrm{C}}=12m_{\mathrm{H}}$. This good agreement between the computer simulation data and the result obtained from relationship (22) indicates that the system is ergodic (the average over the ensemble
of systems is equal to the average over the trajectory of one system), on the one hand, and seems to be rather unexpected, on the other hand. Actually, relationship (22) was derived within the approximation that the
moments of inertia of atoms are constant during the evolution (i.e., under the assumption that changes in the shape and sizes of the cluster are very small), whereas the relative changes in the bond lengths upon vibrations
of the cubane can be rather large and reach $\sim10 \%$ at $T_d\sim 1000$ K. It is interesting to note that, as follows from relationship (22), the constraint on the angular momentum can lead to a nonuniform distribution of the kinetic energy even in clusters composed of atoms with identical masses.

We also simulated the dynamics of the C$_8$H$_8$ cubane with allowance made for the cubane interaction with atoms of a buffer gas, which had a fixed temperature and played the role of a thermal reservoir. As was expected, the velocity distribution of hydrogen and carbon atoms after thermalization of the cluster became Maxwellian irrespective of the initial conditions. The
difference between the kinetic energies of the hydrogen and carbon subsystems was absent.

In conclusion, it should be noted that the nonuniformity of the kinetic energy distribution in small thermally insulated clusters composed of atoms with different masses can be significant and this factor should be taken into account when analyzing rapid processes associated with the decay of clusters and their interaction with each other.

\newpage
\includegraphics[width=\hsize,height=\hsize]{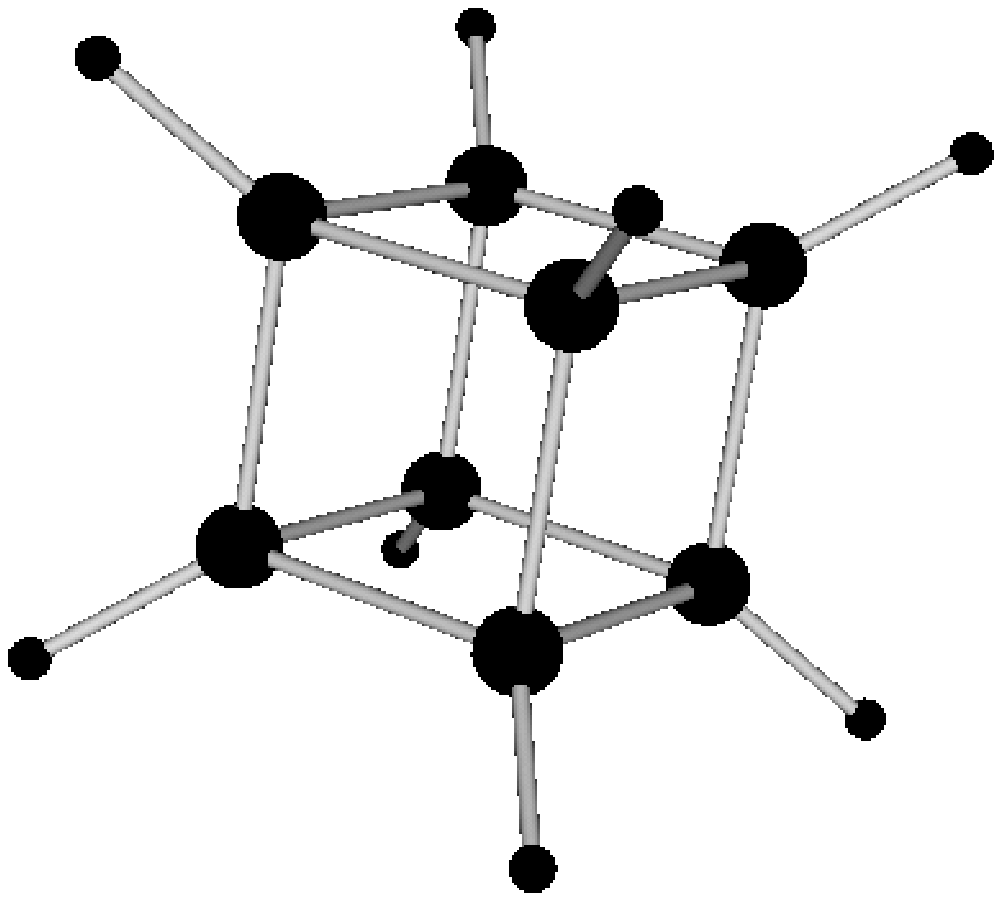}
\vskip 20mm
Fig. 1. Structure of the C$_8$H$_8$ cubane.

\newpage

\includegraphics[width=\hsize,height=\hsize]{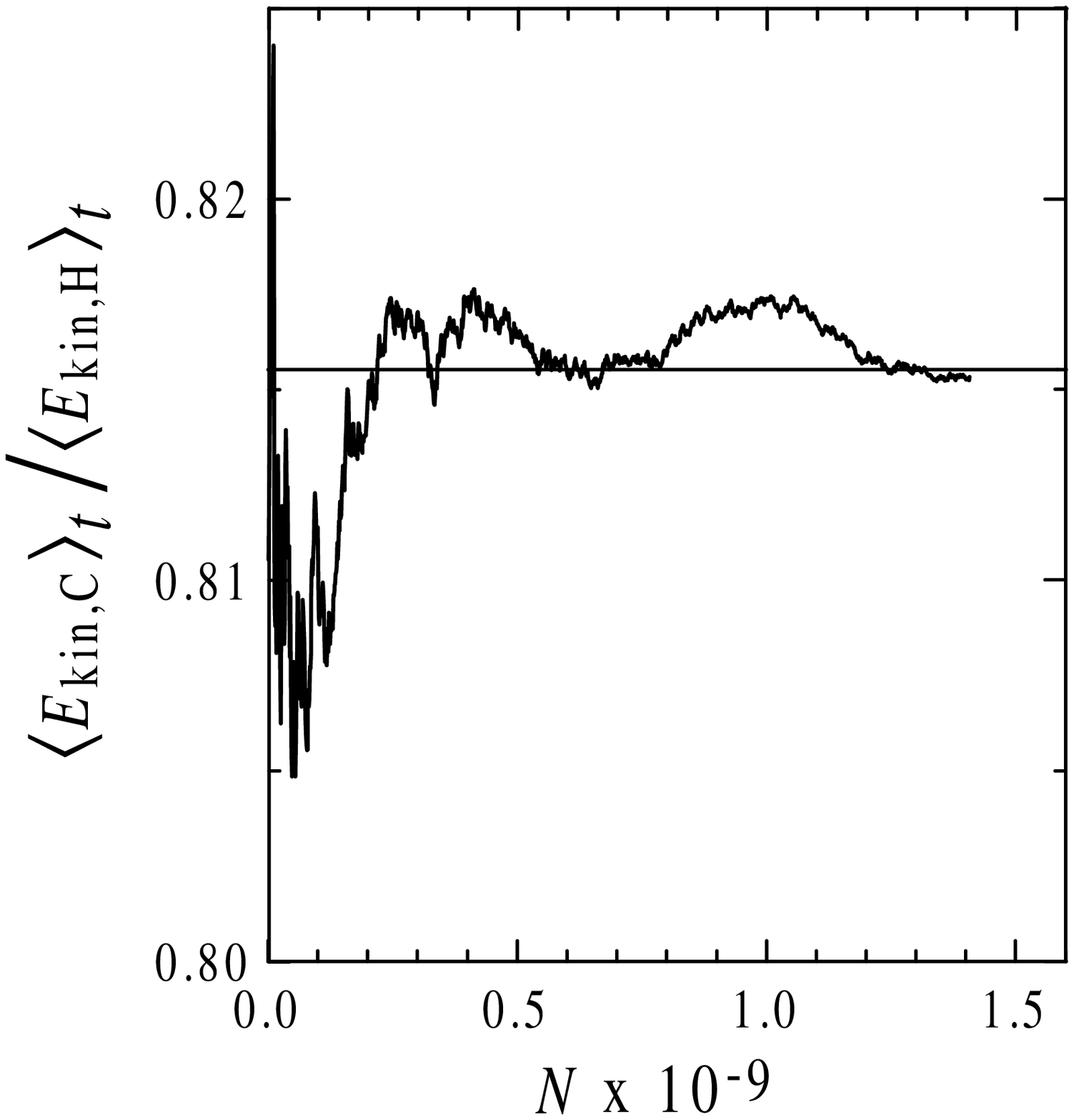}
\vskip 20mm
Fig. 2a. Dependences of the ratio
$\langle E_{kin,\mathrm{C}}\rangle_t/\langle E_{kin,\mathrm{H}}\rangle_t$ between the evolution-time-averaged kinetic energies of the carbon and hydrogen subsystems in the C$_8$H$_8$ cubane on the number of molecular dynamics steps $N$. The horizontal line corresponds to the value
$\langle E_{kin,\mathrm{C}}\rangle_{NVE{\bf PJ}}/\langle
E_{kin,\mathrm{H}}\rangle_{NVE{\bf PJ}}=0.8155$ calculated from relationship (22). Dynamical temperature is $T_d=$ 887 K.

\newpage
\vskip 20mm
\includegraphics[width=\hsize,height=\hsize]{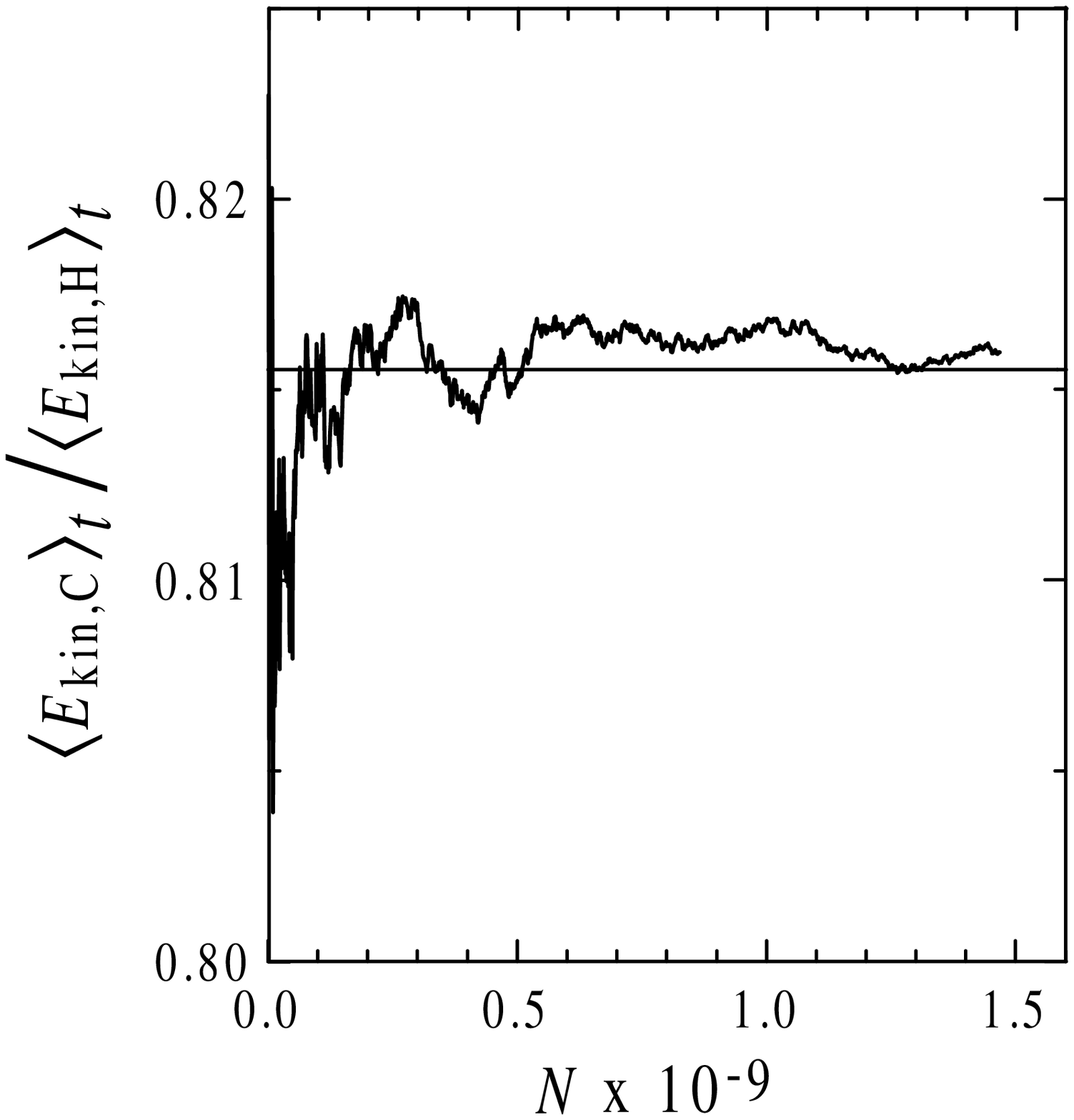}
Fig. 2b. Dependences of the ratio
$\langle E_{kin,\mathrm{C}}\rangle_t/\langle E_{kin,\mathrm{H}}\rangle_t$ between the evolution-time-averaged kinetic energies of the carbon and hydrogen subsystems in the C$_8$H$_8$ cubane on the number of molecular dynamics steps $N$. The horizontal line corresponds to the value
$\langle E_{kin,\mathrm{C}}\rangle_{NVE{\bf PJ}}/\langle
E_{kin,\mathrm{H}}\rangle_{NVE{\bf PJ}}=0.8155$ calculated from relationship (22). Dynamical temperature is $T_d=$ 984 K.

\end{document}